\documentclass[runningheads]{llncs}

\usepackage{amssymb,epsf,psfig}
\newcounter{nbr@exemples}
\newenvironment{exemple}[1]{\begin{example}\textbf{~(#1)}\begin{quote}\begin{small}}{\end{small}\end{quote}\end{example}}
\newenvironment{exemple-continu}[1]{\vspace*{0.3cm}\noindent\textit{Example~#1.}\textbf{~(followed)}\begin{quote}\begin{small}}{\end{small}\end{quote}}
\def\coins{\textsc{coins}}
\def\eg{\emph{eg.}\ }
\def\ie{\emph{i.e.}\ }
\def\csp{\textsc{csp}}

\begin{document}
\setcounter{page}{31}
\title{{COINS}: a constraint-based interactive solving system}
\titlerunning{{COINS}: a constraint-based interactive solving system}
\author{Samir Ouis\inst{1} \and Narendra Jussien\inst{1} \and Patrice Boizumault\inst{2}}
\authorrunning{S. Ouis, N. Jussien, P. Boizumault}
\institute{\'Ecole des Mines de Nantes \\
  4, rue Alfred Kastler -- BP 20722 \\
  F-44307  Nantes Cedex 3 -- France\\
  \email{\{souis,jussien\}@emn.fr} \\
\and
  GREYC, CNRS UMR 6072 \\ Universit\'e de Caen, Campus 2, \\
  F-14032 Caen Cedex -- France\\
  \email{boizu@info.unicaen.fr}}

\maketitle

\addtocounter{footnote}{1}
\footnotetext{In Alexandre Tessier (Ed), proceedings of the 12th International Workshop on Logic Programming Environments (WLPE 2002), July 2002, Copenhagen, Denmark.\\Proceedings of WLPE 2002: \texttt{http://xxx.lanl.gov/html/cs/0207052} (CoRR)}

\begin{abstract}
This paper describes the \coins\ (COnstraint-based INteractive
Solving) system: a conflict-based constraint solver. It helps
understanding inconsistencies, simulates constraint additions
and/or retractions (without any propagation), determines if a given
constraint belongs to a conflict and provides diagnosis tools (\eg
why variable $v$ cannot take value $val$). \coins\ also uses
user-friendly representation of conflicts and explanations.
\end{abstract}

\section{Introduction}

Constraint programming  has been proved extremely successful for
modelling and solving  combinatorial search problems appearing  in
fields such as scheduling  resource allocation and design. Several
languages and systems such as \textsc{chip} \cite{aggoun-chip},
\textsc{choco} \cite{laburthe-choco}, \textsc{gnuProlog}
\cite{diaz-gnuProlog}, \textsc{Ilog solver} \cite{reference01ilog}
have been developed and widely spread. But these systems are
helpless when the constraint network to solve has no solution.
Indeed, only a \texttt{no solution} message is sent to the user
who is left alone to find~:  why the problem has no solution;
which constraint to relax in order to restore the coherence; etc.

These questions yield two different problems: \emph{explaining}
inconsistency and \emph{restoring} consistency. Several
theoretical answers have been provided to address those questions:
\textsc{QuickXPlain} \cite{junker-quickxplain} computes
conflicts for configuration problems,
\cite{bessiere-dynamic} and \cite{debruyne-dynamic} introduce
tools to dynamically remove constraints, \textsc{PaLM}
\cite{jussien-palm} uses conflicts to address those issues and
define new search algorithms, etc.

User interaction requires user-friendly and interactive tools. In
this paper, we advocate for the use of \emph{$k$-relevant}
explanations \cite{bayardo-complexity} to provide the \coins\
(COnstraint-based INteractive Solving) system.

\coins\ helps the user understand inconsistency, simulate
constraint additions and/or retractions (without any propagation),
determine if a given constraint belongs to a conflict and provide
diagnosis tools (\eg why variable $v$ cannot take value $val$).
\coins\ is based upon the use of conflict sets (\emph{a.k.a.}
nogood \cite{schiex-nogood}), explanations
\cite{jussien-e-constraints} and their user-friendly
representation \cite{jussien-user-friendly}.

This paper is organized as follows: we review the definition and
generation of conflicts and explanations within constraint
programming in section~\ref{explanation}. Then, we introduce
$k$-relevance-bounded explanations (section~\ref{sec-k-relevance}) and give an
illustrative example (section~\ref{example}). Before illustrating
the use of $k$-relevant explanations in the \coins\ system
(section~\ref{exploiting}) we present a natural way to provide
user-friendly explanations (section~\ref{user-friendly}). Finally,
we give a short overview of our implementation.

\section{Conflict and explanations for constraint programming}\label{explanation}

A \emph{Constraint Satisfaction Problem} (\csp) is defined by a
set of variables $V = \{v_1, v_2, \ldots, v_n \}$, their
respective value domains $D_1, D_2, \ldots, D_n$ and a set of
constraints $C = \{ c_1, c_2, \ldots, c_m \}$. A solution of the
\csp\ is an assignment of values to all the variables such that
all constraints in $C$ are satisfied. We denote by
$\mathbf{sol}(V,C)$ the set of solutions of the \csp\ $(V,C)$.

In the following, we consider variables domains as unary
constraints. Moreover, the classical enumeration mechanism that is
used to explore the search space is handled as a series of
constraints additions (value assignments) and retractions
(backtracks). Those particular constraints are called
\emph{decision constraints}. This rather unusual consideration
allow the easy generalization of the concepts that are presented
in this paper to any kind of decision constraints (not only
assignments but also precedence constraints between tasks for
scheduling problems or splitting constraints in numeric \csp, etc.
).

\subsection{Definitions}\label{definitions}

Let us consider a constraints system whose current state (\ie the
original constraint and the set of decisions made so far) is
contradictory. A \textbf{conflict set} (\emph{a.k.a.}
\textbf{nogood} \cite{schiex-nogood}) is a subset of the current
constraints system of the problem that, left alone, leads to a
contradiction (no feasible solution contains a nogood). A conflict
divides into two parts\footnote{Notice that some special cases may
arise. If $k<1$, the considered problem is proved as
over-constrained. Some constraints need to get relaxed. If $C' =
\emptyset$, the set of decisions that has been taken so far is in
itself contradictory. This can happen only if no propagation is
done after a decision has been made.}: a subset of the original
set of constraints ($C' \subset C$ in equation~\ref{eq-nogood})
and a subset of decision constraints introduced so far in the
search (here $dc_1, \ldots, dc_k$).

\begin{equation} \label{eq-nogood}
   \mathbf{sol}\left(V,\left(C' \wedge dc_1 \wedge ... \wedge dc_k
   \right)\right) = \emptyset
\end{equation}

An operational viewpoint of conflict sets can be made explicit by
rewriting equation~\ref{eq-nogood} the following way:
\begin{equation} \label{eq-explanation}
   C' \wedge \left(\bigwedge_{i \in [1..k] \setminus j}dc_i\right)\rightarrow \neg dc_{j}
\end{equation}

Let us consider $dc_j: v_j = a$ in the previous formula.
Equation~\ref{eq-explanation} leads to the following result
($s(v)$ is the value of variable $v$ in the solution $s$):

\begin{equation} \label{eq-explanation-rewritten}
   \forall s \in \mathbf{sol}\left(V,C' \wedge \left(\bigwedge_{i \in [1..k] \setminus j}dc_i
   \right)\right), s(v_j) \neq a
\end{equation}

The left hand side of the implication in
equation~\ref{eq-explanation} is called an \textbf{eliminating
explanation} (explanation for short) because it justifies the
removal of value $a$ from the domain $d(v)$ of variable $v$. It is
noted: $expl(v \neq a)$.

Explanations can be combined to provide new ones. Let us suppose
that $dc_1 \lor \ldots \lor dc_j$ is the set of all possible
choices for a given decision (set of possible values, set of
possible sequences, etc.).  If a set of explanations $C'_1
\rightarrow \neg dc_1$, ...,  $C'_j \rightarrow \neg dc_j$ exists,
a new conflict can be derived: $C'_1 \wedge \ldots \wedge C'_j$.
This new conflict provides more information than each of the old
ones.

For example, a conflict can be computed from the empty domain of a
variable $v$ (using explanations for each of the removed values):
\begin{equation} \label{eq-contradiction}
   \bigwedge_{a \in d(v)} expl(v \neq a)
\end{equation}


\subsection{Storing explanations: $k$-relevance-bounded learning}
There generally exists several explanations for the removal of a
given value. Several different approaches were introduced to
handle that multiplicity. \emph{Dependency Directed
Backtracking}\cite{stallman-ddb} records all encountered
explanations. The major inconvenience of this approach is its
exponential space complexity. Indeed,  the number of recorded
explanations increases in a monotonous way. Various algorithms
only keep a \emph{single} explanation: \emph{Dynamic Backtracking}
\cite{ginsberg-dynamic} and its improvements (\emph{MAC-DBT}
\cite{jussien-macdbt-cp}, \emph{Generalized Dynamic Backtracking}
\cite{bliek-gpb}, \emph{Partial-order Dynamic Backtracking}
\cite{ginsberg-gsat}) and \emph{Conflict-directed BackJumping}
\cite{prosser-maccbj}. The idea is to forget (erase) explanations
which are not valid any more considering the current set of
decision constraints. Space complexity therefore remains
polynomial while ensuring the completeness of the algorithms.
Unfortunately, this idea is not really compatible with debugging:
only one explanation is kept and past information are completely
lost.

Instead of recording only one explanation, a more interesting idea is to
keep information as long as a given criterion is verified:

\begin{itemize}
\item Time-bounded criterion: explanations are forgotten after a
given time. This criteria is similar to \emph{tabu} list
management in \emph{tabu} search \cite{glover-tabu}.

\item Size-bounded criterion:  \cite{schiex-nogood-orig} have used
a criteria defined in \cite{dechter90}. This criteria keeps only
the explanations with a size lower or equal to a given value $n$.
This criteria limits the spatial complexity, but may forget really
interesting nogoods.

\item Relevance-bounded criterion: explanations are kept if they are
not \emph{too far} from the current set of decision constraints. This
concept (called $k$-relevance) has been introduced in
\cite{bayardo-complexity} and focus explanations/conflict
management to what is important: relevance \emph{wrt} the current
situation.
\end{itemize}

Time and size-bounded  recording do have a controllable space
complexity. This is also the case for $k$-relevance learning
(\emph{cf.} section~\ref{sec-k-relevance}). As we shall see, our
tools are meant for the debugging and the dynamic analysis of programs:
the space occupation overhead is well worth it.

\section{$k$-relevance-bounded explanations}
\label{sec-k-relevance}

While solving a constraint problem, the current state of calculus can be
described with two sets of constraints: \textbf{$R$ the set of relaxed
constraints} (decisions that have been undone during search,
constraints that have been explicitly relaxed by the user, etc.)
and $A$ the set of active constraints (the current constraint
store). $\langle A,R\rangle$ is called a \emph{configuration}.
Following \cite{bayardo-complexity}, we can now define a
$k$-relevant explanation as:

\begin{definition}{$k$-relevant explanation (\cite{bayardo-complexity})}\\
Let $\langle A,R\rangle$ a configuration. An explanation $e$ is
said to be \textbf{$k$-relevant} if it contains at most $k-1$
relaxed constraints, \ie $|e \cap R|<k$.
\end{definition}

In $k$-relevance-bounded learning, only $k$-relevant explanations
are kept during search. Hence, several different explanations may
be kept for a given value removal. Thus $expl(v \neq a)$
will not contain any more a single explanation but the set of
$k$-relevant explanations recorded for the removal of value $a$ from the domain
$d(v)$ of variable $v$.

%
%

\subsection{Managing $k$-relevant explanations}

\subsubsection{Computing $k$-relevant explanations}
$k$-relevant explanations, as regular explanations
\cite{jussien-e-constraints} can be computed during propagation.
However, some issues arise (see example~\ref{ex-issues}).

\begin{exemple}{Example for explanation computation}\label{example-computation}
\label{ex-issues} Let us consider two variables $v_1$ and $v_2$.
Let us assume that value $a$ from $v_1$ is only supported by value
$b$ from $v_2$ in constraint $c$. Let finally assume that $b$ is
removed from $v_2$ (a set of explanations being: $\{ \{c_1, c_2
\}, \{c_1,c_3\}, \{c_4, c_5\}\}$).
This removal needs to be
propagated.

But, which explanation one should choose to compute the
explanation of the value removal $v_1 \neq a$ ? Do we have to
consider all the possibilities  $ \{c, c_1, c_2
\}$, $\{c, c_1,c_3\}$ or $\{c, c_4, c_5\}$? Only one ?
\end{exemple}

As values are removed only once, we can focus on one particular
explanation: the first one. The explanation computed at removal
time is called the \emph{main} one. That explanation will be used
to compute forthcoming explanations. Moreover, this explanation is
exactly the one that would have been computed by a classical
approach (see section~\ref{definitions}).

\begin{exemple-continu}{\ref{ex-issues}}
Let us suppose that the \emph{main} explanation for the removal of
value $b$ from $v_2$ is $\{c_1, c_2\}$.

Thus, the removal $v_1 \neq a$ will be justified by $\{c, c_1,
c_2\} $.
\end{exemple-continu}

\subsubsection{Evolution of the $k$-relevant explanations}

We need to maintain the relevance information attached to stored
explanations upon constraint additions and retractions. In both
ways, the relevance of explanations may vary. The idea is to keep
track of these variations and to forget explanations as soon as
they become irrelevant. We organize  the contents of the
$k$-relevant explanations according to the  relevance of its
explanations. More precisely, all  $k$-relevant explanations for a
given removal $expl(v \neq a)$ are partitioned into $k$ subsets,
\ie\ $expl(v \neq a) = \cup_{i \in [0..k-1]} expl(v \neq a,i)$.
Subset $expl(v \neq a, i)$ contains the explanations having $i$
relaxed constraints.

\begin{exemple}{Updating $2$-relevant explanations}
\label{example-2-relevance}

Consider example~\ref{example-computation}. We assume that no
other explanation has been found for the removal $v_1\not=
a$. The $2$-relevant explanations for this removal are~:
$expl(v_1 \not= a,0) = \{\{c, c_1, c_2\}, \{c_2,c_4\}\}$,
$expl(v_1 \not= a,1) = \{\{c_1, c_3\},\{c_2, c_3\}\}$.

Notice that constraint $c_3$ is relaxed. That is why both $\{c_1,
c_3\}$ and $\{c_2, c_3\}$ are in the $expl(v_1 \not= a,1)$.

\begin{itemize}
\item If constraint $c_1$ is relaxed: explanation $\{c, c_1, c_2\}$
jumps from $expl(v_1 \not= a, 0)$ to $expl(v_1 \not= a, 1)$;
explanation $\{c_1,c_3\}$ is forgotten and thus removed from
$expl(v_1 \not= a, 1)$ (we have $k=2$). The new set of
$2$-relevant explanations is therefore: $expl(v_1 \not= a,0) = \{
\{c_2,c_4\}\}$, $expl(v_1 \not= a,1) = \{\{c, c_1, c_2\}, \{c_2,
c_3\} \}$.
\item If constraint $c_3$ is reactivated (from the original set of explanations): explanations
$\{c_1, c_3\}$ and $\{c_2, c_3\}$ become valid and the new set of
$2$-relevant explanations is therefore: $expl(v_1 \not= a,0) = \{
\{c, c_1, c_2\}, \{c_2,c_4\}, \{c_1, c_3\},\{c_2, c_3\}\}$,
$expl(v_1 \not= a,1) = \{ \}$.
\end{itemize}
\end{exemple}

\subsubsection{Computing conflicts}
The same dilemma that we encountered when computing explanations
appears when computing conflicts. Indeed, when a contradiction is
identified (a domain of a variable becomes empty), we saw that
equation~\ref{eq-contradiction} computes a conflict. However,
there may exist several explanations for each considered value
removal. Contrarily to the explanation computation process, we
chose here to provide all possible explanations (limiting
ourselves to valid explanations \ie $expl(v \neq a,0)$) for all $i
\in [0..k-1]$. The resulting   number of valid conflicts is:
\begin{equation} \label{eq-all-conflicts}
\quad \prod_{a\in d(v)} |expl(v\not= a, 0)| \quad
\mathtt{conflicts}
\end{equation}

\begin{exemple}{The possible nogoods}
Let us consider example~\ref{example-2-relevance} but with a
different $k$. We assume that values $a$ and $b$ from $v_1$ are
have been removed with the following explanations: $expl(v_1 \not=
a,0) = \{ \{c, c_1, c_2\}, \{c_2,c_4\}\}$ and $expl(v_1 \not= b,0)
= \{ \{c_1, c_5\}, \{c_2,c_5\}\}$.

Applying equation~\ref{eq-all-conflicts} leads to the following
nogoods:
\begin{itemize}
\item $\{c, c_1, c_2\} \cup \{c_1, c_5\}$
\item $\{c, c_1, c_2\} \cup \{c_2, c_5\}$
\item $\{ c_2, c_4\} \cup \{c_1, c_5\}$
\item $\{c_2, c_4\} \cup \{c_2, c_5\}$
\end{itemize}

Here, the \emph{main} explanation for $v_1 \not= a$ is $\{c, c_1,
c_2\}$ and $\{c_1, c_5\}$  is the one for $v_1 \not= b$.
\end{exemple}


\subsection{$1$-relevance \emph{vs.} classical approaches}

All classical approaches (\emph{Dynamic Backtracking} or
\emph{MAC-DBT}) forget explanations as they become invalid. A
$1$-relevant learning technique will obviously proceed the same
way. However, it differs from classical approaches by  the number
of recorded explanations by removal. Indeed, during resolution,
one may come across an explanation for an already performed
removal. Instead of not taking it into account, $1$-relevance will
keep that secondary information\footnote{It will be used to compute conflict. The
\emph{main} explanation will still be the only used to compute
subsequent explanations.}.

Furthermore, all classical approaches take into account only one
conflict. This conflict is computed following to
equation~\ref{eq-contradiction}. Our approach may deal with a
\emph{set} of conflicts (see equation~\ref{eq-all-conflicts}).
Nevertheless that particular explanation management has a
computational and spatial cost.

\subsection{Complexity issues}\label{complexity}

To compute the complexity of our approach, let us consider a \csp\
defined upon $n$ discrete variables with maximum domain size $d$
upon which are posted $e$ constraints. If we only keep a single
explanation per value removal, there will be at most $n \times d$
explanations of maximal size $e + n$ \ie all the constraints from
the problem ($e$) and the decision constraints  ($n$). Thus the
complexity of the classical approach is $O((e+n) \times n\times d)
$.

However, as far as the $k$-relevance approach is concerned, an
explanation can contain up to $k-1$ relaxed constraints, the
maximal size of an explanation being $n + e + k - 1$. The maximum
number of explanations for a given value removal is bounded by the
maximum number of non included subsets in a set.
The worst case is: $\left(\begin{array}{c}  { e+n+k-1 } \\
{(e+n+k-1)/2} \end{array}\right)$ subsets of size $(e+n+k-1)/2$.

Therefore, the spatial complexity for storing $k$-relevance
explanations is in: $$O\left(n\times d \times  \left(\begin{array}{c} { e+n+k-1 } \\
{(e+n+k-1)/2}  \end{array}\right) \times (e+n+k-1)/2\right)$$

\section{An example~: the conference problem} \label{example}
To illustrate the use of the $k$-relevant explanations, we present
the resolution of the  conference problem
\cite{jussien-implementing-lncs}. From now on, we fill focus our
study to $1$-relevance.



\subsection{Presentation of the problem}\label{problem-conference}

Michael, Peter and Alan are organizing a two-day seminar for
writing a report on their work. In order to be efficient, Peter
and Alan need to present their work to Michael and Michael needs
to present his work to Alan and Peter (actually Peter and Alan
work in the same lab). Those presentations are scheduled for a
whole half-day each. Michael wants to known what Peter and Alan
have done before presenting his own work. Moreover, Michael would
prefer not to come the afternoon of the second day because he has
got a very long ride home. Finally, Michael would really prefer
not to present his work to Peter and Alan at the same time.

\subsection{A constraint model for the conference problem} \label{conf-model}
A constraint model for that problem is described as follows~: let
$Ma, Mp, Am, Pm$ the variables representing four presentations
($M$ and $m$ are respectively for Michael as a speaker and as an
auditor). There domain will be $[1,2,3,4]$ ($1$ is for the morning
of the first day and $4$ for the afternoon of the second day).
Several constraints are contained in the problem: implicit
constraints regarding the organization of presentations and the
constraints expressed by Michael.

The implicit constraints can be stated:
\begin{itemize}
\item A speaker cannot be an auditor in the same half-day. This
constraint is modelled as: $c_1: Ma \neq Am$, $c_2: Mp \neq Pm$,
$c_3: Ma \neq Pm$ and $c_4: Mp \neq Am$.
\item No one can attend two presentations at the same time. This
is modelled as $c_5: Am \neq Pm$.
\end{itemize}

Michael constraints can be modelled:
\begin{itemize}
\item Michael wants to speak after Peter and Alan: $c_6: Ma > Am$,
$c_7: Ma > Pm$, $c_8: Mp > Am$ and $c_9: Mp > Pm$.
\item Michael does not want to come on the fourth half-day: $c_{10}: Ma \neq 4$,
$c_{11}: Mp \neq 4$, $c_{12}: Am \neq 4$ and $c_{13}: Pm \neq 4$.
\item Michael does not want to present to Peter and Alan at the
same time: $c_{14}: Ma \neq Mp$.
\end{itemize}

\subsection{Using classical approaches}
Table~\ref{DomainePalmenum} shows the resulting explanations for
both approaches (classical and $1$-relevant) after adding
constraints from $C_1$ to $C_6$.
%
%
The column associated to the classical approach contains only a
single explanation by removal, opposed to the $1$-relevant column.
The domain of the variable $P_m$ is empty. We deduce that our
problem is over-constrained. According to the
equation~\ref{eq-contradiction} of the section~\ref{explanation},
we obtain the conflict $\{C_1, C_2, C_3, C_4, C_5, C_6\}$.

\begin{table}[htpb]
  \begin{center}
    \leavevmode

    \begin{tabular}{|ccccc|} \hline
      Variable & Value & Explanation  & $1$-relevance & present~?  \\ \hline
       $P_m$ & $1$ & $\{C_1,C_2,C_4,C_6\}$ & $\{C_1,C_2,C_4,C_6\},\{C_1,C_4,C_6\} $ & no \\
      $P_m$ & $2$ & $\{C_5,C_6\}    $&$\{C_5,C_6\}    $ & no \\
      $P_m$ & $3$ & $\{C_5,C_6\}  $ & $\{C_5,C_6\} $ &no \\
      $P_m$ & $4$ & $\{C_3\}  $ &$\{C_3\},\{C_5\},\{C_5,C_6\}  $ &no \\ \hline
      $A_m $ & $1$ & $\emptyset   $ & $\emptyset   $ &\textbf{yes} \\
      $A_m $ & $2$ & $\{C_4,C_6\}   $ & $\{C_4,C_6\}   $  &no \\
      $A_m $ & $3$ & $\{C_4,C_6\}   $ & $\{C_4,C_6\}   $  &no \\
      $A_m $ & $4$ & $\{C_2\}   $ & $\{C_2\},\{C_4\},\{C_4,C_6\}   $ &no \\ \hline
      $M_p$ & $1$ & $\{C_4\}   $ &  $\{C_4\},\{C_5\},\{C_6\}   $  &no \\
      $M_p$ & $2$ & $\emptyset   $ & $\emptyset   $  &\textbf{yes}      \\
      $M_p$ & $3$ & $\{C_6\}  $ &$\{C_6\}  $ &no \\
      $M_p$ & $4$ & $\{C_6\}   $ & $\{C_6\}  $ &no \\ \hline
      $M_a $ & $1$ & $\{C_2\}   $ &$\{C_2\},\{C_3\}   $ &no \\
      $M_a $ & $2$ & $\emptyset   $ &  $\emptyset   $ &\textbf{yes} \\
      $M_a $ & $3$ & $\emptyset   $ &  $\emptyset   $ &\textbf{yes} \\
      $M_a $ & $4$ & $\emptyset   $ &  $\emptyset   $ &\textbf{yes} \\ \hline
    \end{tabular}

    \caption{Domains after the introduction of constraints}
    \label{DomainePalmenum}
  \end{center}
\end{table}

\subsection{Solving using $1$-relevant explanations}

Table~\ref{DomaineKrelevance} presents the $1$-relevant
 explanations associated to every removal after we have
removed the redundant explanations  like $\{C_5,C_6\}$ for the
removal $P_m \not= 4$. But as the second approach proposes several
explanations, we can deduce several conflicts. In our case, we
obtain two conflicts~: $\{C_1, C_3, C_4, C_5, C_6\}$ and $\{C_1,
C_4, C_5, C_6\}$.

The second conflict is more precise since  it is included in the
first one. There is a quite important difference between the
conflict provided by the first approach which contains all the
constraints that do not help the user and the conflicts provided
by the  $1$-relevant approach.

\begin{table}[[htpb]
  \begin{center}
    \leavevmode

    \begin{tabular}{|ccccc|} \hline
      Variable & Value & Explanation & $1$-relevance & present~?  \\ \hline
      $P_m$ & $1$ & $\{C_1,C_2,C_4,C_6\}$ &$\{C_1,C_4,C_6\} $ & no \\
      $P_m$ & $2$ &$\{C_5,C_6\}    $ &$\{C_5,C_6\}    $ & no \\
      $P_m$ & $3$ & $\{C_5,C_6\}  $ &$\{C_5,C_6\}  $ & no \\
      $P_m$ & $4$ & $\{C_3\}  $ &$\{C_3\},\{C_5\}  $ & no \\ \hline
      $A_m $ & $1$ &$\emptyset   $ &$\emptyset   $ & \textbf{yes} \\
      $A_m $ & $2$ &$\{C_4,C_6\}   $ &$\{C_4,C_6\}   $ & no \\
      $A_m $ & $3$ & $\{C_4,C_6\}   $ &$\{C_4,C_6\}   $ & no \\
      $A_m $ & $4$ & $\{C_2\}   $  &$\{C_2\},\{C_4\}   $ & no \\ \hline
      $M_p$ & $1$ & $\{C_4\}   $  &$\{C_4\},\{C_5\},\{C_6\}   $ &  no \\
      $M_p$ & $2$ & $\emptyset   $ &$\emptyset   $ &  \textbf{yes}      \\
      $M_p$ & $3$ & $\{C_6\}  $ &$\{C_6\}  $ & no \\
      $M_p$ & $4$ & $\{C_6\}  $ &$\{C_6\}   $ & no \\ \hline
      $M_a $ & $1$ & $\{C_2\}   $  &$\{C_2\},\{C_3\}   $ & no \\
      $M_a $ & $2$ & $\emptyset   $ &$\emptyset   $ &  \textbf{yes} \\
      $M_a $ & $3$ & $\emptyset   $ &$\emptyset   $ &  \textbf{yes} \\
      $M_a $ & $4$ & $\emptyset   $ &$\emptyset   $ &  \textbf{yes} \\ \hline
    \end{tabular}

    \caption{Final set of explanations  }
    \label{DomaineKrelevance}
  \end{center}
\end{table}

\subsection{User interaction}

As we can see in our example, conflicts and explanations are sets
of low-level constraints. Only a specialist can understand and
correctly interpret the provided information. In order to design
user interaction tools, we therefore need to provide
\emph{translation} tools to make accessible to any user the
low-level constraints. In the next section, we introduce
user-friendly explanations to address that issue.

\section{User-friendly explanations} \label{user-friendly}

In this section, we introduce the notion of user-friendly
explanations \cite{jussien-user-friendly}. We provide a set of
tools to address the accessibility issue of conflicts and
explanations made of low-level constraints. 

A solution to that issue will be obtained thanks to the developer
of the application. When developing an application, such an expert
needs to \emph{translate} the problem from the high level
representation (the user's comprehension of the problem) to the
low-level representation (the actual constraints in the system).
We note this translation a \texttt{user} $\rightarrow$
\texttt{system} translation.

For user-friendly explanations, we need the other way
\emph{translation}: from the low-level constraints (solver
adapted) to the user comprehensible constraints (higher level of
abstraction). We note this translation a \texttt{system}
$\rightarrow$ \texttt{user} translation. That translation is
usually not explicitly coded in the system. Asking a developer to
provide such a translator while coding would be quite strange for
him. We chose to automatize that translation in a transparent way.

\subsection{Hierarchical representation of problems}
\begin{figure}[hbtp]
  \begin{center}
  ~\hspace{1.8cm}
 \epsfbox{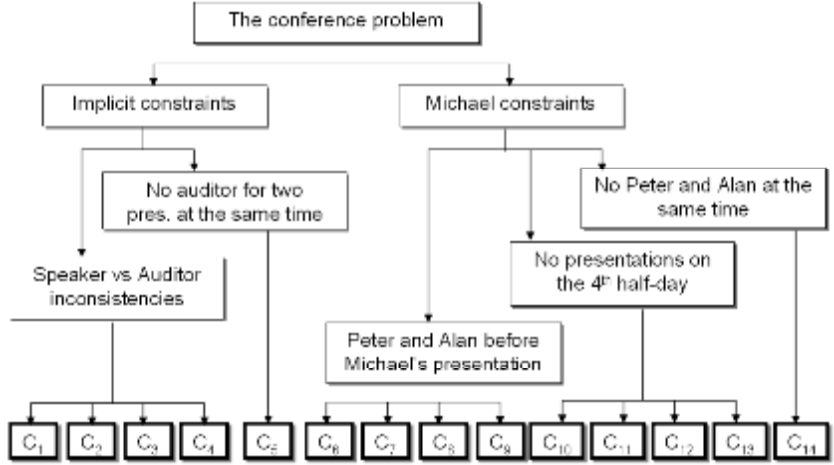}
   ~\vspace{-23cm}
  \end{center}
\caption{An hierarchical view of the conference problem} \label{fig-conference}
\end{figure}

This idea relies on a single hypothesis: all aspects of a
constraint-based application can be represented in a hierarchical
way. Indeed, any object appearing in a constraint problem is
attached to at most a single \emph{father-}object. Note that a
given \emph{father-}object may have several
\emph{children-}objects. For example, figure~\ref{fig-conference}
gives a graphical representation of the problem defined in
section~\ref{problem-conference}.

\subsection{Building a \texttt{system} $\rightarrow$ \texttt{user} translator}

While developing a constraint application, the user only
needs to explicitly state the underlying hierarchy of his problem.
Only the leaves of this structure, namely the low-level
constraints, can be used by the constraint solver.

The leaves may be too low-level for a typical user of the final
application. However, he can understand higher levels in the
hierarchy. What allows the hierarchy hypothesis is to build with
no effort for the developer a hierarchical representation of the
problem. Once built, this representation can be used to interact
with any user through user-friendly explanations. Such
explanations are provided using procedures converting low-level
constraints into user understandable pieces of the hierarchy.
Those procedures are completely problem independent and may be
provided within the constraint solver.

The user perception of a given problem can be seen as a \emph{cut}
in the hierarchical view of the considered problem. For example,
let suppose that the user is Michael who does not want to deal
with implicit constraints. Although he does understand his own
wishes. Therefore, his view of the problem would be: \fbox{The
conf. problem}, \fbox{P\&A before}, \fbox{Not 4$^{th}$ $1/2$ day}
and \fbox{P\&A not same time}.

Our example has no solution, one explanation for that situation
provided by a classical approach is: $\{c_1, c_2, c_3, c_4, c_5,
c_6\}$.

Michael looks at the explanation from his point of view. We
therefore need to project the concrete explanation onto his
representation. The projection consists in projecting the
constraints $\{c_1, c_2, c_3, c_4, c_5, c_6\}$ from bottom to the
top of the hierarchy representing the problem (see
figure~\ref{fig-conference}) until a user understandable box is
reached.

For example, the projection of the constraints $\{c_1, c_2, c_3,
c_4\} $ gives in first step the \fbox{Speaker vs. Auditor} box.
Unfortunately, this box is not understandable by Michael. In this
case, the projection continues to the father box: \fbox {Implicit
constraints}. Once again, this box is not understandable by
Michael and the projection gets to \fbox {The conf. problem}.
Finally, we reached \fbox{The conf. problem} box which Michael can
deal with. The projection of $c_5$ gives the same box through
\fbox{Auditor vs. 2 pers.} and \fbox{Implicit constraints}. For
$c_6$, the projection is easier because the first reached box is
user understandable.

The final projection gives: \fbox{The conf. problem} and  \fbox{P\&A
before}.  He can ask to discard one box. For example, the box
\emph{Michael wants to speak after Peter and Alan}.


\section{Exploiting $k$-relevant explanations} \label{exploiting}

$k$-relevance provides more interesting explanations and allows to
obtain a better diagnosis. In this section, we present five
concrete situations which the user is frequently confronted in the
case of a failure. We show how  $k$-relevant explanations allow
to better understand and to analyze this failure and to quickly
simulate various scenarios.


\subsection{Giving more precise explanations}



If we have two explanations for the same removal $e_1$ and $e_2$
such as $e_1 \subsetneq e_2$, then we know that the constraints
which  belong to $e_2 \setminus e_1$ are not responsible for the
removal. We say that $e_1$ is more precise than $e_2$.

%

Consequently, constraints belonging to $e_2 - e_1$ are not
responsible of the incoherence if they do not appear in the other
removals. $k$-relevance is, of course, not the panacea (see
constraint $C_6$ which is not responsible for the removal  $P_m
\not = 4$ but intervenes in the removal  $Pm \not = 1$  -- it
appears in the conflict). But, the multiplicity of explanations
leads to more precise explanations and conflicts.

\subsection{Does a constraint belong to a conflict?}
\label{sec-belonging}


An interesting issue when debugging is to know wether a given
constraint belongs  to a conflict or not. $k$-relevant explanations
help answering that question.

Let suppose that the cause of incoherence is the variable $A_m$
(see table~\ref{DomaineAm}). As there is a failure (the domain of
variable $A_m$ is empty), the user wants to know  if
\textit{constraint $C_5$ belongs to a conflict} by referring to
the table~\ref{DomaineAm}. Based  on the classical approach, the
only conflict will be $\{C_2,C_3,C_4,C_6\}$.  Indeed, The answer
will be negative ($C_5 \not\in \{C_2, C_3, C_4, C_6 \} $). While
our $1$-relevant approach provides 8 conflicts  and indicates that
the constraint $C_5$ is strongly responsible of the incoherence.

\begin{table}[h]
  \begin{center}
    \leavevmode

    \begin{tabular}{|ccccc|} \hline
      Variable & Value & Explanation & $1$-relevance &not deleted~?  \\ \hline

      $A_m $ & $1$ & $\{C_3\}$ & $\{C_3\},\{C_5\}$ & no \\
      $A_m $ & $2$ & $\{C_4,C_6\}   $ &$\{C_4,C_6\},\{C_5\}   $ & no \\
      $A_m $ & $3$ & $\{C_4,C_6\}   $ &$\{C_4,C_6\},\{C_5\}   $ & no \\
      $A_m $ & $4$ & $\{C_2\}       $ & $\{C_2\},\{C_4\}   $ & no \\ \hline

   \end{tabular}

    \caption{ New state of the variable $A_m$}
    \label{DomaineAm}
  \end{center}
\end{table}

\subsection{Simulating constraint relaxation}
Determining if a given constraint belongs to a conflict or not may
lead to spend a lot of time by relaxing each suspected constraint.
For that reason, we propose a tool which allows to simulate a
relaxation (without any propagation) only by updating the
$k$-relevant explanations.

For example, let suppose that the user suspects that constraint
$C_3$ belongs to a conflict. The constraint-checking tool (see
section~\ref{sec-belonging}) confirms it. The relaxation of this
constraint will put back all the values $a$ such as $C_3 \in
expl(A_m \not= a)$. According to table~\ref{DomaineAm}, the
constraint $C_3$ is partly responsible for the removal $A_m \not=
1$. The classical approach would have put back the value $1$ in
the domain of $A_m$ and launched the propagation phase.
Unfortunately, the problem is always over-constrained because  the
removal $Am \not= 1$ is justified by the constraint $C_5$ and the
domain of $A_m$ becomes empty again.

$1$-relevant explanations allow us to know that the relaxation of
the constraint $C_3$ will lead  towards an another failure  due to
the removal $A_m \not= 1$ that will be justified by another
explanation:  $\{C_5\}$. Thus, our tool is able to indicate to the
user if a relaxation of a suspected constraint will lead to
another \emph{immediate} failure.

\subsection{Simulating constraint addition}

To solve a dynamic problem, a simple  execution from  scratch
is too expensive for every modification introduced by the user.
Some proposed tools allowing to solve the problem from the current
solution do not allow to know if this modification (precisely the
addition of a previously relaxed constraint) will lead towards a
failure.


It is helpful to take advantage of the information accumulated
during the resolution of the previous problem to avoid adding
constraints leading towards an \emph{immediate} failure. For this
reason, we have proposed a tool simulating the re-introduction of
a relaxed constraint without any propagation. This tool informs
the user if the addition of a relaxed constraint leads towards a
failure or not. So we can avoid   reintroducing such constraints.

For example, let suppose now that the user has removed constraints
$\{c_3, c_5\}$ to put back value~1 in the domain of $A_m$. The new
domains (along with the associated explanations) are reported in
table~\ref{NewDomaineAm}. Let suppose that our user wants to put
back the previously relaxed constraint $c_3$. A naive approach
consists in propagating the constraint from the current situation
(it leads to a contradiction). However, $2$-relevant explanations
can be used to simulate this constraint addition by updating the
relevance status of associated explanations. Some of them may
become valid ($\{c_3\}$ in our example) and therefore remove some
values (1 from $A_m$ here). Here, with no propagation at all,
using $k$-relevant explanation could have helped the user by
telling him that adding constraint $c_3$ would have lead to a
contradiction.

\begin{table}[hbtp]
  \begin{center}
    \leavevmode

    \begin{tabular}{|cccccc|} \hline
      Variable & Value & Explanation & $1$-relevance & $2$-relevance & present~?  \\ \hline

      $A_m $ & $1$ & & & $\{C_3\},\{C_5\}$ & yes \\
      $A_m $ & $2$ & $\{C_4,C_6\}   $ &$\{C_4,C_6\},\{C_5\}   $ & & no \\
      $A_m $ & $3$ & $\{C_4,C_6\}   $ &$\{C_4,C_6\},\{C_5\}   $ & & no \\
      $A_m $ & $4$ & $\{C_2\}       $ & $\{C_2\},\{C_4\}   $ & & no \\ \hline

   \end{tabular}

    \caption{ New state of the variable $A_m$ after relaxation of $c_3$ and $c_5$}
    \label{NewDomaineAm}
  \end{center}
\end{table}



\subsection{Providing error diagnosis}
Imagine now that after some relaxations, the user wants to know
why the variable $M_p$ cannot take the value $1$? The classical
approach provides the explanation $\{C_6\}$.  While the
$1$-relevant approach provides the set of explanations: $\{
\{C_4\}, \{C_5\}, \{C_6\}\}$.

We notice that $1$-relevance provides a richer diagnosis than the
classical approach.

\section{Implementation} \label{implementation}

\textbf{COINS} has been  implemented in
\texttt{choco}\footnote{\texttt{choco} is an open source
constraint engine developed as the kernel of the OCRE project. The
OCRE project (its name standing for \emph{Outil Contraintes pour
la Recherche et l'Enseignement}, \ie, \emph{Constraint tool for
Research and Education}) aims at building free Constraint
Programming tools that anyone in the Constraint Programming and
Constraint Reasoning community can use. For more information see
www.choco-constraints.net.} \cite{laburthe-choco} using   the
\texttt{PaLM} system \cite{jussien-palm}.  \texttt{choco} allows~:
to propagate the constraints, to manage domains as well as other
filtering algorithms, local search, etc. Experiments show that
when  $k$ increases, the performance of $k$-relevance decreases.
In the other hand, for $k=1$ and $k = 2$, we obtain the same
temporal performances as \textbf{mac-dbt}\cite{jussien-macdbt-cp}.
For $k = 1$ and $k = 2$, the time lost to manage the $k$-relevant
explanations  is compensated  by avoiding future failures. For $k
\geq 3$, $k$-relevance loses its advantages. More precisely, for
such a $k$, we are losing time managing explanations that will
seldom let us avoid future failures. Especially, we shall update
explanations which will never become valid (\ie $1$-relevant).

\section{Conclusion}
In this paper, we have introduced the foundations of several
interactive tools which are of great help for a user to solve an
over-constrained problem. We have shown the effectiveness of these
$k$-relevance-based tools. $k$-relevance keeps several
explanations by removal and forgets them once they become
irrelevant.

We have shown the contribution of our $k$-relevance-based tools
compared to  classical approach.  $k$-relevance
provides more precise explanations; gives some general information
that cannot be accessible within a classical framework:
$k$-relevance allows the simulating of constraint
retraction/addition and so provides richer diagnosis tools.

Our current work includes designing algorithms which compute the
\emph{best} conflict from all the explanations. We investigate
adding user-based comparators \cite{borning-constraint-89} to our
tools in order to provide automatic comparison of solutions.
Also, we try to decrease the space complexity by managing differently the explanations.

\end{document}